\documentclass[11pt]{cernrep}
\usepackage{graphicx}
\usepackage{cite,./mcite,amsmath,amssymb}
\newcommand{\beeq}{\begin{eqnarray}}
\newcommand{\eeeq}{\end{eqnarray}}
\newcommand{\be}{\begin{equation}}
\newcommand{\ee}{\end{equation}}
\newcommand{\bea}{\begin{array}}
\newcommand{\eea}{\end{array}}

\def\alfasbar{\overline{\alpha}_s}

\def\ubo{\boldsymbol{u}}
\def\vbo{\boldsymbol{v}}
\def\wbo{\boldsymbol{w}}

\begin{document}

\title{Dipole models and parton saturation in $ep$ scattering 
}

\author{L.~Motyka$^{1,2}$, K.~Golec-Biernat$^{3,4}$ and G. Watt$^5$}
\institute{
$^1$ II Institute for Theoretical Physics, Luruper Chaussee 149, 22761 Hamburg, Germany\\  
$^2$ Institute of Physics, Jagellonian University, Reymonta 4, 30-059 Krak\'{o}w, Poland\\
$^3$ Institute of Nuclear Physics, Polish Academy of Sciences, Krak\'{o}w, Poland\\ 
$^4$ Institute of Physics, University of Rzesz\'{o}w, Rzesz\'{o}w, Poland \\
$^5$ Department of Physics \& Astronomy, University College London, WC1E 6BT, UK}

\maketitle

\begin{abstract}
In this contribution we briefly review the current status of the dipole models and parton saturation on the basis of results presented at the HERA--LHC workshops in the years 2006--2008. The problem of foundations of the dipole models is addressed within the QCD formalism. Some limitations of the models and open problems are pointed out. Furthermore, we review and compare the currently used dipole models and summarise the applications to describe various sets of HERA data. Finally we outline some of the theoretical approaches to the problem of multiple scattering and saturation.
\end{abstract}

\section{Introduction}

Dipole models \cite{Nikolaev:1990ja,*Nikolaev:1991et,Mueller:1993rr,*Mueller:1994jq,GolecBiernat:2008} represent a QCD motivated framework that has been successfully applied to describe a variety of gluon mediated scattering cross sections at high energies. In particular, they provide a transparent and intuitive picture of scattering processes. Their main strength is a combination of universality, simplicity and efficiency.  The dipole models are capable of simultaneously describing all $F_2$, $F_L$ and heavy quark production $ep$ data at small~$x$, the inclusive diffractive data, the bulk of measurements for exclusive diffractive vector meson production, deeply virtual Compton scattering (DVCS), and even nuclear shadowing \cite{GolecBiernat:1998js,*GolecBiernat:1999qd,Bartels:2002cj,*GolecBiernat:2006ba,Kowalski:2003hm,Forshaw:2004vv,*Forshaw:2006np,*Shaw:2007,Iancu:2003ge,Soyez:2007kg,Marquet:2007qa,Kowalski:2006hc,Watt:2007nr,*Watt:2008,Kowalski:2007rw,*Kowalski:2008sa}. This unified description is achieved using only a few parameters with a transparent physical meaning, such as the normalisation of the gluon distribution at a low scale, the quark mass or the proton size. At the same time, the dipole models provide a phenomenological insight into important aspects of high energy scattering, like the relative importance of multiple scattering or higher twist contributions. This importance may be quantified in terms of a {\em saturation scale}, $Q_{S}$, the scale of the process at which the unitarity corrections become large~\cite{GolecBiernat:1998js,*GolecBiernat:1999qd}. Up to now, the dipole models applied to HERA data offer one of the most convincing arguments for the dependence of this scale on the scattering energy and provide one of the best quantitative estimates  of the saturation scale~\cite{GolecBiernat:1998js,*GolecBiernat:1999qd,Bartels:2002cj,Kowalski:2003hm,Kowalski:2006hc,Watt:2007nr}. This shows the complementarity of dipole models to the rigorous framework of collinear factorisation, within which the description of multiple scattering, although possible in principle, is quite inefficient. It is not only very demanding from the technical side (for instance, even the basis of twist-four operators is not fully understood yet), but it would also require introducing a set of new unknown functions parameterising the expectation values of higher twist operators at the low (input) scale. In dipole models this problem is bypassed by simply fitting the (implicitly) resummed multiple scattering cross section together with the nonperturbative contribution with constraints imposed by the unitarity of the scattering matrix.

\section{Foundations and limitations of dipole models}

Let us consider a $\,2\to 2\,$ scattering amplitude of 
$\;i + p \,\to\, f + p\,$, 
where the strongly interacting projectile $i$ hits a hadronic target 
$p$ and undergoes a transition to a state $f$, while the target scatters 
elastically. At HERA the projectile is always a virtual photon, $\gamma^*$, 
with a four-momentum $q$ and virtuality $q^2 = -Q^2$, and the target is a 
proton, with initial momentum $p$ and final momentum $p^\prime$. The final states considered are virtual and 
real photon states, vector meson states and diffractive states.  
The states $i$~and~$f$ carry a typical scale $\bar Q^2$; for 
$i=f=\gamma^*(Q^2)$, $\bar Q^2 = Q^2$. The invariant collision energy 
$s=(p+q)^2$ is assumed to be large, $s \gg \bar Q^2$ and $s \gg |t|$, 
where $t=(p-p^\prime)^2$~is the momentum transfer.
We shall also use the variable $x = \bar Q^2 / s$, that reduces to 
the Bjorken~$x$ for the case of deeply inelastic scattering (DIS).

\begin{figure}     \begin{center}
      \includegraphics[width=0.35\textwidth]{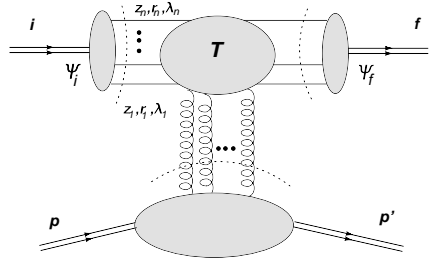}
    \end{center}
  \caption{High energy scattering in the dipole representation.} 
  \label{fig:dipgen}
\end{figure}

The key idea behind dipole models is a separation (factorisation) of a high energy scattering amplitude, $\mathcal{A}^{i\,p\rightarrow f\,p}$, into an initial ($\Psi_i$) and final ($\Psi_f$) state wave function of the projectile $i$ and the outgoing state $f$, and a (diagonal) universal scattering amplitude of a multi-parton Fock state, ${\cal F}_n$, off a target $p$; see Fig.~\ref{fig:dipgen}.
The scattering operator, $T$, is assumed to be diagonal in the basis of states that consist of a definite number of partons, $n$, with fixed longitudinal momentum fractions, $z_k$ ($k=1,\ldots,n$), of the projectile, definite helicities, $\lambda_k$, and transverse positions, $\boldsymbol{r}_k$.  One may write symbolically (see e.g.\cite{Bartels:2007aa}):
\begin{equation} \label{eq:genfact}
  \mathcal{A}^{i\,p\rightarrow f\,p}
= \; 
\sum_{\mathrm{n,{\cal F}_n,\{\lambda_k\}}}\,
\int\![\mathrm{d}^{2n}\boldsymbol{r}_k]\int\![\mathrm{d^n}z_k] \;
\Psi_{f}^{*}(n,\{z_k,\boldsymbol{r}_k, \lambda_k\}) 
\;T({\cal F}_n)\; 
\Psi_{i}^{}(n,\{z_k,\boldsymbol{r}_k,\lambda_k\}).
\end{equation}
In most practical applications one takes into account only the lowest
Fock states, composed of a quark--antiquark ($q\bar q$) pair and, 
possibly, one additional gluon ($q\bar q g$). In the limit of a large 
number of colours, $N_c \to \infty$, flavourless scattering states, 
$i$ and $f$, may be represented as a collection of colour 
dipoles~\cite{Mueller:1993rr,*Mueller:1994jq}.  
For the simplest case of $q\bar q$ scattering, the intermediate state
${\cal F}_2$ is defined by the quark and antiquark helicities, 
the longitudinal momentum fraction, $z$, of the projectile carried by 
the quark, the dipole vector, 
$\boldsymbol{r}= \boldsymbol{r}_2 - \boldsymbol{r}_1$,
and the impact parameter vector, 
$\boldsymbol{b} = z\boldsymbol{r}_1 + (1-z)\boldsymbol{r}_2$. 
It is convenient to define the imaginary part of the dipole scattering amplitude (assuming 
independence of the azimuthal angles), 
$\mathcal{N}(x,\boldsymbol{r},\boldsymbol{b}) \equiv \mathrm{Im}\,T({\cal F}_2),\,$
and the $b$-dependent dipole--target cross-section
\be
\frac{\mathrm{d}\sigma_{q\bar q}}{\mathrm{d}^2\boldsymbol{b}} = 
2\;\mathcal{N}(x,r,b).
\ee

The picture encoded in (\ref{eq:genfact}) may be motivated within 
perturbative QCD. 
In the high energy limit of QCD~\cite{Lepage:1980fj,Gribov:1984tu}, 
the dominant contribution to scattering amplitudes comes from vector 
boson (gluon) exchanges, that lead to cross-sections constant with energy 
(modulo quantum corrections that may generate an additional enhancement). 
For each spin-1/2 fermion (quark) exchange in the $t$-channel 
the amplitude is power suppressed by a factor of $1/s^{1/2}$. 
In consequence, the high energy scattering amplitude may be factorised into
the amplitude describing slow (in the target frame) gluon fields 
and the amplitude of fast parton fields of the projectile moving in the 
gluon field of the target. This is, in fact, the basic 
assumption of the $k_T$- (high energy) 
factorisation~\cite{Gribov:1984tu,Catani:1990eg,*Catani:1994sq}. 
In the high energy limit, the vertex describing the coupling of the 
fast $s$-channel parton (quark or gluon) to a gluon exchanged in 
the $t$-channel is {\em eikonal}: the large light-cone component of 
the longitudinal parton momentum and the parton helicity are conserved.
Also, up to subleading terms in the collision energy, the fast parton 
does not change its transverse position in the scattering process. 
These properties of high energy amplitudes in QCD were used to derive the dipole model for hard processes. 
In more detail, the scattering amplitudes in the dipole model follow from the QCD scattering amplitudes obtained within the $k_T$-factorisation scheme, in the high energy limit and at the leading logarithmic (LL) $\ln (1/x)$ 
approximation~\cite{Nikolaev:1990ja,*Nikolaev:1991et}.

The fact that the QCD dipole model follows from the $k_T$-factorisation approximation implies that the model, up to subleading terms in $1/s$, is also consistent with the leading order (LO) collinear approximation~\cite{Catani:1990eg,*Catani:1994sq}. In addition, as in the case of the $k_T$-factorisation framework, the dipole model incorporates an exact treatment of the quark transverse momentum in the box diagram. These kinematic effects, when analysed within the collinear approximations, manifest themselves as higher order corrections to the coefficient functions~\cite{Catani:1990eg,*Catani:1994sq}. Although the implicit resummation of the collinear higher order terms in the dipole model is only partial, it should still be viewed as an improvement of the LO collinear approximation.

Practical use of dipole models is not restricted to hard processes, where precise predictions can be obtained within the collinear factorisation framework. On the contrary, one of the most successful applications of the dipole model (the saturation model~\cite{GolecBiernat:1998js,*GolecBiernat:1999qd}) provides an efficient and simple description of the transition from the perturbative single scattering regime (the colour transparency regime) to the multiple scattering regime as a function of the process scale and scattering energy (or $\bar Q^2$ and $x$). In this transition region scattering amplitudes are expected to receive contributions both of the nonperturbative nature and from perturbatively calculable multiple scattering effects. The nonperturbative effects in high energy scattering are currently not computable with theoretical methods and have to be modelled.
The multiple scattering effects enter the scattering amplitudes e.g.\ as higher twist contributions~\cite{Bartels:2000hv,*Motyka:2008}\footnote{Multiple scattering effects that occur at low scales are absorbed into the input gluon density at the initial scale.}, that are suppressed by inverse  powers of the hard scale, $\bar Q^2$, and additional powers of~$\alpha_s$. Nevertheless, the higher twist effects may be quite sizable at small~$x$ and at moderately large $\bar Q^2$~\cite{Bartels:2000hv,*Motyka:2008}. This originates from a rapid growth of the multi-gluon density with decreasing~$x$: assuming the large $N_c$~limit, the $n$-gluon density evolves approximately as the single gluon density to power~$n$~\cite{Bukhvostov:1985rn,Bartels:1993ke}. Thus, at decreasing~$x$ the multiple scattering effects are increasingly enhanced and may eventually become competitive with the single scattering contribution.

Thus far we discussed the dipole model from the perspective of perturbative
QCD. An interesting attempt to provide foundations of the model in a 
general (i.e.\ non-perturbative) framework was recently put 
forward~\cite{Ewerz:2004vf,*Ewerz:2006vd,vonManteuffel:2007}. 
The scattering amplitudes were written in terms of 
skeleton diagrams and the QCD path-integral. Approximations and assumptions 
necessary to recover the dipole model amplitudes were identified. To a large 
extent the conclusions from that analysis confirm those obtained 
within the perturbative framework: the dipole model accuracy is not 
theoretically guaranteed when higher twist and higher order corrections 
are large. An interesting point raised in Refs.~\cite{Ewerz:2007md,vonManteuffel:2007} is the dependence of the dipole cross section, $\sigma_{q\bar{q}}$, 
on the dipole--target collision energy, $\sqrt{s}$. 
In most models one assumes that $\sigma_{q\bar{q}}$  depends on $s$ through 
$x=\bar Q^2 /s$.
The scale, however, is part of the wave functions and it is not obvious that 
the dipole cross section should depend on $\bar Q^2$ rather than on the dipole variables, like e.g.\ the dipole scale, $1/r^2$. Interestingly, assuming the dependence of $\sigma_{q\bar{q}}$ on a combined variable$\,s\,r^2\,$ was shown to create some tension between the HERA data on $F_2$ and $F_L$ and the dipole model, irrespective of the detailed functional form of $\sigma_{q\bar{q}}$. Some insight may be gained from inspecting the issue in the $k_T$-factorisation approach. Then, the energy dependence enters through $x_g$~of the gluon, that essentially depends on the external state virtuality, the scattered quarks' transverse momenta and the distribution of the quark longitudinal momentum. So, the proposed replacement of $\bar Q^2$ by $1/r^2$ might be somewhat oversimplified. On the other hand, within the LL$(1/x)$ approximation the standard choice of $x_g \simeq \bar Q^2 / s$ is justified. 
To sum up, the choice of the optimal dimensionless variable that would carry 
the energy dependence of the dipole cross-section remains an open and 
interesting problem.

\section{Phenomenology of dipole models}

\begin{figure}
  \begin{minipage}{0.5\textwidth}
    (a)
    \begin{center}
      \includegraphics[width=0.8\textwidth]{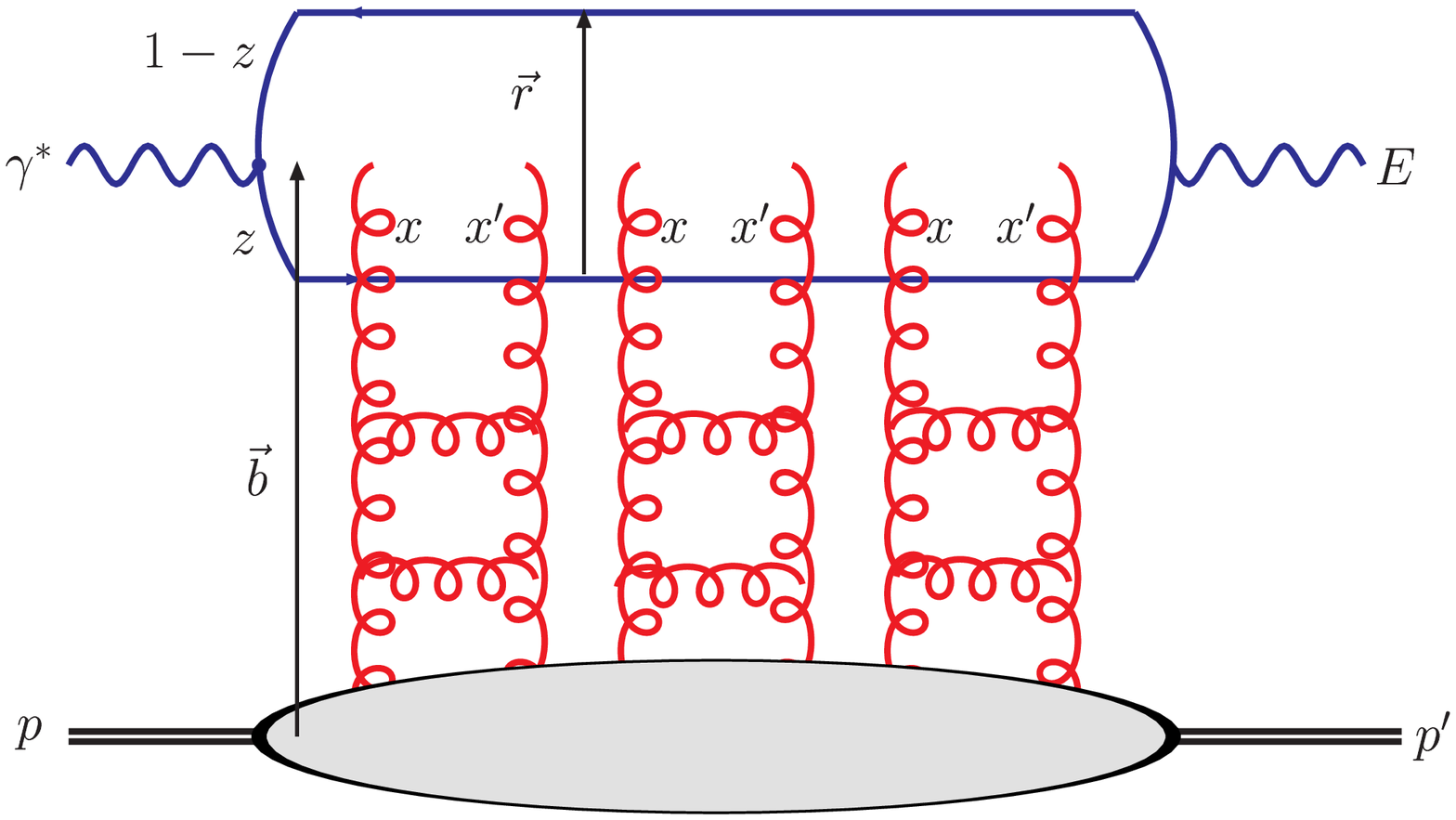}
    \end{center}
  \end{minipage}%
  \begin{minipage}{0.5\textwidth}
    (b)
    \begin{center}
      \includegraphics[width=0.8\textwidth]{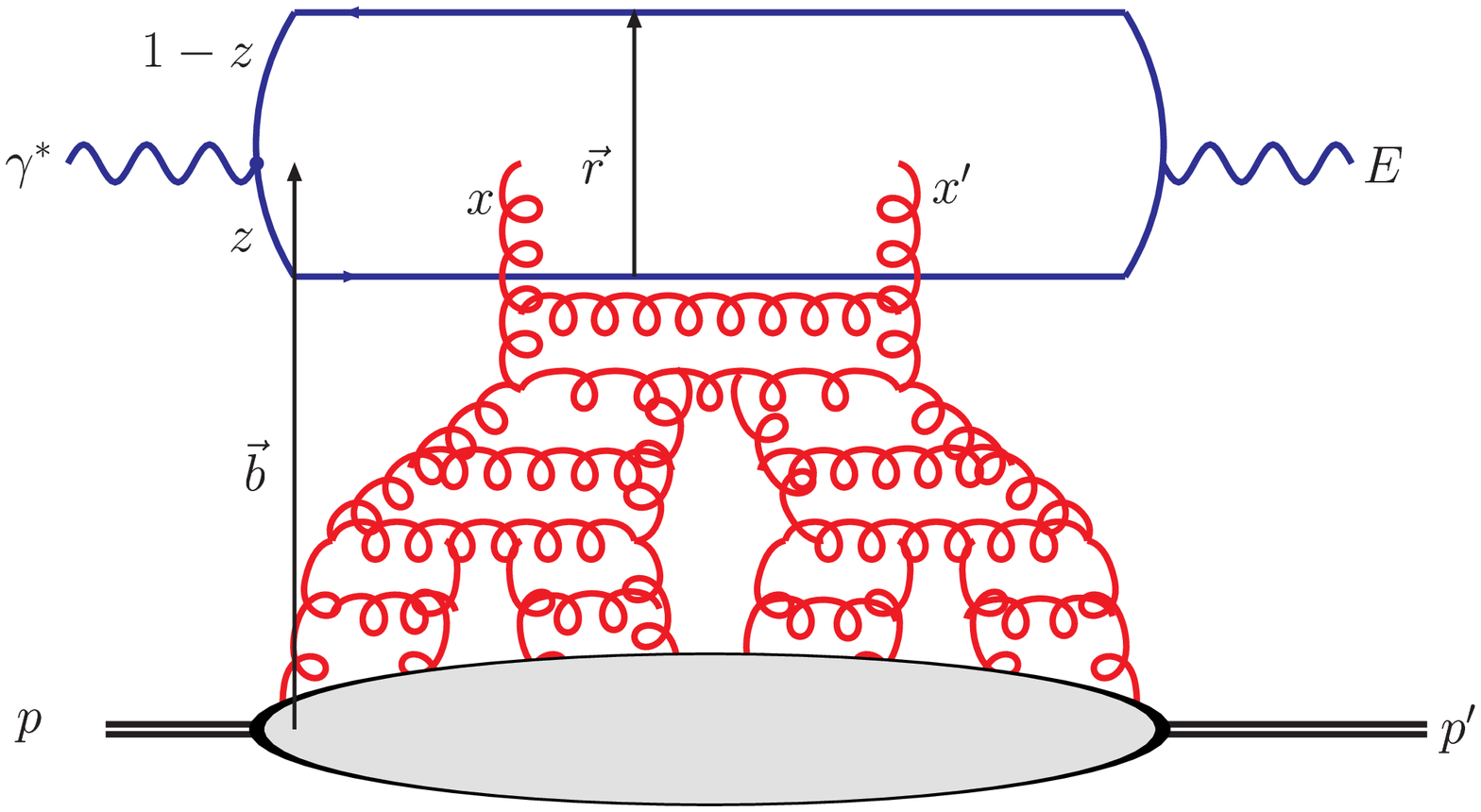}
    \end{center}
  \end{minipage}
  \caption{The $\gamma^*p$ scattering amplitude with unitarisation achieved via (a) \emph{eikonal} diagrams or (b) \emph{fan} diagrams.  For exclusive diffractive processes, such as vector meson production ($E=V=\Upsilon,J/\psi,\phi,\rho$) or DVCS ($E=\gamma$), we have $x^\prime\ll x\ll 1$ and $t=(p-p^\prime)^2$.  For inclusive DIS, we have $E=\gamma^*$, $x=x^\prime\ll 1$ and $p=p^\prime$.}
  \label{fig:diagrams}
\end{figure}

Implementations of multiple scattering in colour dipole models are based on
two main approaches, that adopt different approximations. 
The Glauber--Mueller (GM) \emph{eikonal} 
approach~\cite{Glauber:1955,*Mueller:1989st}
is used in the family of models that evolved from the Golec-Biernat--W\"usthoff
 (GBW) model \cite{GolecBiernat:1998js,*GolecBiernat:1999qd}. 
One assumes in this approach that multiple colour dipole scatters are 
independent of each other, see Fig.~\ref{fig:diagrams}a.
This assumption may be supported (although it was not yet explicitly 
derived) with properties of the collinear evolution of quasi-partonic 
operators describing the multi-gluon density in the proton, and in the large 
$N_c$~limit~\cite{Bukhvostov:1985rn,Bartels:1993ke,Bartels:2000hv,*Motyka:2008}. Assuming in addition a factorised $b$-dependence of the gluon distribution, one postulates the dipole--proton scattering amplitude of the form:
\begin{equation} \label{eq:b-Sat}
  \mathcal{N}(x,r,b) = 1-\exp\left(-\frac{\pi^2}{2N_c}r^2\alpha_s(\mu^2)\,xg(x,\mu^2)\,T(b)\right),
\end{equation}
where the scale $\mu^2 = C/r^2 + \mu_0^2$ with $\mu_0 \sim 1$ GeV. HERA data on exclusive vector meson production imply a Gaussian form of the proton shape in the transverse plane, $T(b)$, with $\sqrt{\langle b^2\rangle} = 0.56$~fm. The corresponding quantity determined from the proton charge radius (0.87~fm) is somewhat larger, $\sqrt{\langle b^2\rangle} = 0.66$~fm, implying that gluons are more concentrated in the centre of the proton than quarks.  The form \eqref{eq:b-Sat} is denoted by the ``b-Sat'' model \cite{Kowalski:2003hm,Kowalski:2006hc}.  It can be considered to be an improvement on a previous model \cite{Bartels:2002cj,*GolecBiernat:2006ba} where $T(b)\propto \Theta(R_p-b)$ was assumed, and also on the original GBW model \cite{GolecBiernat:1998js,*GolecBiernat:1999qd} where additionally the scale dependence of the gluon distribution was neglected, that is, $xg(x,\mu^2)\propto x^{-\lambda}$ was assumed for a fixed power $\lambda\sim 0.3$.  Note that in the GBW model large saturation effects were needed to get from the hard Pomeron behaviour ($\sim r^2\,x^{-0.3}$) at small dipole sizes to soft Pomeron behaviour ($\sim x^{-0.1}$) at large dipole sizes.  On the other hand, in Refs.~\cite{Bartels:2002cj,*GolecBiernat:2006ba,Kowalski:2003hm,Kowalski:2006hc} this transition can alternatively be achieved with DGLAP evolution, therefore saturation effects are correspondingly smaller.

In the alternative approach one exploits solutions of the Balitsky--Kovchegov 
(BK) equation~\cite{Balitsky:1995ub,*Kovchegov:1999yj,*Kovchegov:1999ua}. 
It was derived for scattering of a small colour dipole off a large nucleus, 
composed of $A$~nucleons. 
The LL~BK equation rigorously resums contributions of BFKL Pomeron \emph{fan} 
diagrams  (Fig.~\ref{fig:diagrams}b), that are leading in $A$, $1/N_c$ and 
in the $\ln 1/x$ approximation (properties of solutions of the next-to-LL~BK
equation are not known yet and so cannot be used in the dipole models). 
A colour glass condensate (CGC) dipole model parameterisation \cite{Iancu:2003ge} was constructed from an approximate solution of the BK equation:
\begin{equation} \label{eq:cgc}
  \mathcal{N}(x,r,b) = T(b)\,\mathcal{N}(x,r) \, = \, \Theta(R_p-b)
  \begin{cases}
    \mathcal{N}_0\left(\frac{rQ_s}{2}\right)^{2\left(\gamma_s+\frac{\ln(2/rQ_s)}{9.9\lambda \ln(1/x)}\right)} & :\quad rQ_s\le 2\\
    1-\mathrm{e}^{-A\ln^2(BrQ_s)} & :\quad rQ_s>2
  \end{cases},
\end{equation}
where $Q_s = (x_0/x)^{\lambda/2}$ is a saturation scale.\footnote{In what follows we shall use $Q_s$ (with a lower-case $s$) to denote the saturation scale defined in a model-dependent way.} The original analysis \cite{Iancu:2003ge} neglected the charm quark contribution to $F_2$. The inclusion of charm was later found \cite{Kowalski:2006hc} to significantly lower the saturation scale when the anomalous dimension $\gamma_s$ was fixed at the LO BFKL value of $0.63$.  By letting $\gamma_s$ go free, a solution was subsequently found with $\gamma_s = 0.74$ which included heavy quarks but had a large saturation scale \cite{Soyez:2007kg}.  (This model has been modified to include a $t$ dependence in the saturation scale allowing the description of exclusive diffractive processes \cite{Marquet:2007qa}.)  However, the HERA data do not show a strong preference for the solution with $\gamma_s = 0.74$, and a secondary solution with $\gamma_s = 0.61$ and a much smaller saturation scale also describes the data well \cite{Watt:2007nr,*Watt:2008}.  The CGC model \eqref{eq:cgc} assumes a factorised $b$ dependence which is not supported by HERA diffractive data, where one finds a significantly non-zero effective Pomeron slope $\alpha_\mathbb{P}^\prime$, indicating correlation between the $b$ and $x$ dependence of the dipole scattering amplitude.  A more realistic impact parameter dependence was included by introducing a Gaussian $b$ dependence into the saturation scale $Q_s$, denoted by the ``b-CGC'' model \cite{Kowalski:2006hc,Watt:2007nr,*Watt:2008}.  It was not possible to obtain a good fit to HERA data with a fixed $\gamma_s = 0.63$ \cite{Kowalski:2006hc}, but on freeing this parameter, a good fit was obtained with a value of $\gamma_s = 0.46$ \cite{Watt:2007nr,*Watt:2008}, close to the value of $\gamma_s \simeq 0.44$ obtained from numerical solution of the BK equation \cite{Boer:2007wf}.  However, the value of $\lambda = 0.119$ obtained from the ``b-CGC'' fit \cite{Watt:2007nr,*Watt:2008} is lower than the perturbatively calculated value of $\lambda\sim 0.3$ \cite{Triantafyllopoulos:2002nz}.

\begin{figure}
  \begin{minipage}{0.5\textwidth}
    (a)\\
    \includegraphics[width=\textwidth]{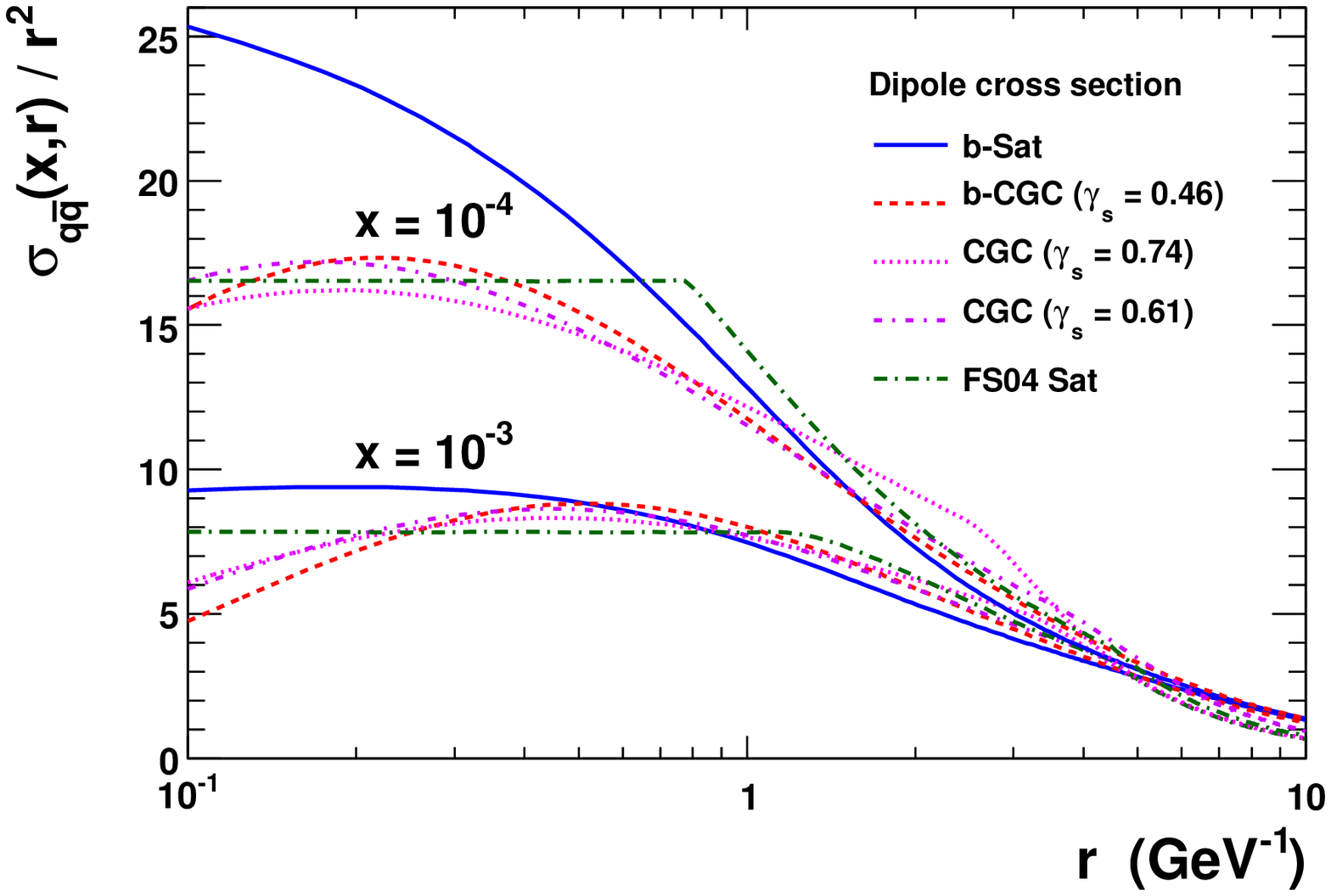}
  \end{minipage}%
  \begin{minipage}{0.5\textwidth}
    (b)\\
    \includegraphics[width=\textwidth]{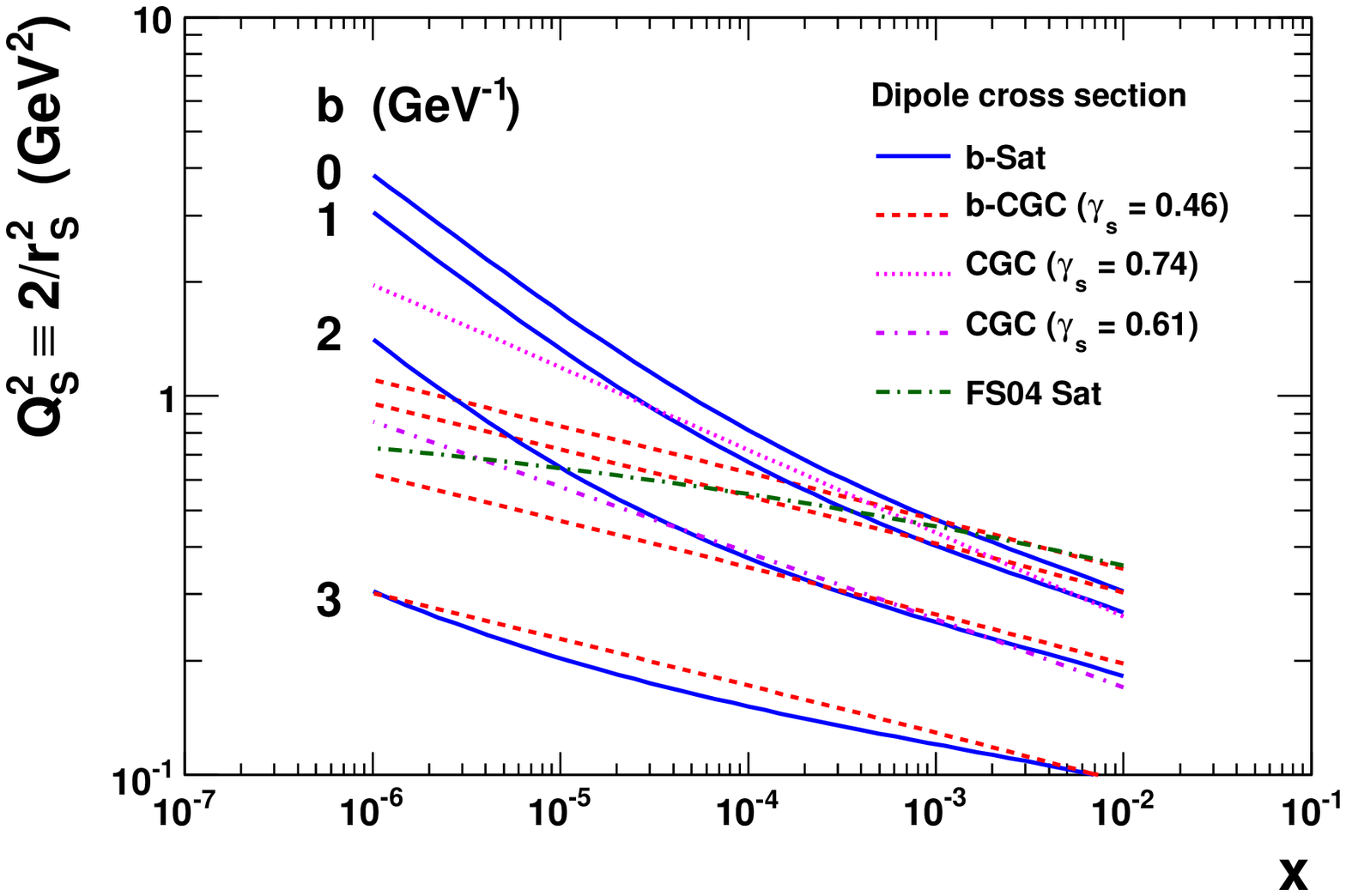}
  \end{minipage}
  \caption{(a) 
The $b$-integrated dipole--proton cross sections divided by $r^2$ and
(b) the saturation scale $Q_S^2\equiv 2/r_S^2$. }
  \label{fig:plots}
\end{figure}

In both the approaches to unitarisation one neglects multi-gluon correlations 
in the target. Thus, the key difference between the eikonal and the BK 
approaches is that in the latter one resums the leading logarithms of $1/x$
while in the former one aims at keeping a reasonable representation of 
leading logarithms of $\bar Q^2$. 
Both dipole model realisations have built in saturation of the black disc 
limit of the colour dipole scattering amplitude.
This means that the absolute value of the $T$-matrix elements tends to unity 
for large dipoles or as $x\to 0$. It is curious that the choice 
of approximation has a striking consequence in how the unitarity 
(the black disc) limit is approached. In the GM case unitarisation happens because of cancellations between contributions of non-saturating 
multiple gluon exchanges, while in the BK case multiple scattering effects are 
contained in the single gluon density that saturates at a certain small value of~$x$. 
These differences in the mechanism of unitarisation do not affect, however, 
the crucial qualitative feature of the dipole cross-section: the transition 
from a power-like growth with decreasing~$x$ in the colour transparency 
regime to a flat (possibly $\sim \ln(1/x)$) behaviour in the black disc limit.
Thus, the necessary modelling of the dipole cross section for large dipole
sizes is strongly constrained.

A third type of parameterisation for the dipole cross section does not assume 
any mechanism for unitarisation. It is a two-component Regge model (FS04 Sat) \cite{Forshaw:2004vv,*Forshaw:2006np,*Shaw:2007}, which uses hard Pomeron behaviour ($\sim r^2\,x^{-0.3}$) for small dipole sizes $r<r_0$ and soft Pomeron behaviour ($\sim x^{-0.1}$) for large dipole sizes $r>r_1$, with linear interpolation between the two regions.  Again, a factorising impact parameter dependence is assumed.  Saturation effects are modelled by allowing $r_0$ to move to lower values with decreasing $x$.  This feature was found to be preferred by the HERA data, whereas the two-component Regge model with a fixed $r_0$ was disfavoured \cite{Forshaw:2004vv,*Forshaw:2006np,*Shaw:2007}.

We compare the dipole model parameterisations in Fig.~\ref{fig:plots}a, where the $b$-integrated dipole cross sections have been divided by the trivial factor $r^2$ in order to emphasise the differences at small $r$.  We restrict attention to dipole model parameterisations which have been shown to give a good fit (with charm quarks included) to recent HERA inclusive structure function data, meaning a $\chi^2$ per data point of $\sim 1$.  This excludes, for example, the original GBW parameterisation \cite{GolecBiernat:1998js,*GolecBiernat:1999qd} and the unsaturated two-component Regge model \cite{Forshaw:2004vv,*Forshaw:2006np,*Shaw:2007}.  All parameterisations shown in Fig.~\ref{fig:plots}a are similar at intermediate dipole sizes where they are most constrained by HERA data.  At very small dipole sizes the b-Sat model deviates from the other parameterisations, as it is the only one which incorporates explicit DGLAP evolution.  The b-Sat model was found to be preferred over the b-CGC model for observables sensitive to relatively small dipole sizes \cite{Watt:2007nr,*Watt:2008}.  There are also differences between the parameterisations in the approach to the unitarity limit at large dipole sizes.  For example, the b-Sat and b-CGC dipole cross sections tend to a constant at large $r$ only for a fixed $b$, but not when integrating over all impact parameters.

In order to compare the magnitude of unitarity corrections between various models it is customary to define a model-independent saturation scale $Q_S^2$, that is, the momentum scale at which the dipole--proton scattering amplitude $\mathcal{N}$ becomes sizable.  There is no unique definition of $Q_S^2$ and various choices are used in the literature.  We define the saturation scale $Q_S^2\equiv 2/r_S^2$, where the saturation radius $r_S$ is the dipole size where the scattering amplitude
\begin{equation} \label{eq:satdef}
  \mathcal{N}(x,r_S[,b]) = 1 - \mathrm{e}^{-\frac{1}{2}} \simeq 0.4,
\end{equation}
chosen to match the corresponding quantity, $Q_s$, in the GBW model \cite{GolecBiernat:1998js,*GolecBiernat:1999qd}.  Note that this ``saturation scale'' is still far from the unitarity limit where $\mathcal{N}=1$.  The model-independent saturation scale $Q_S^2$ is shown in Fig.~\ref{fig:plots}b: it is generally less than $0.5$ GeV$^2$ in the HERA kinematic regime for the most relevant impact parameters $b\sim 2$--$3$ GeV$^{-1}$~\cite{Kowalski:2006hc,Watt:2007nr,*Watt:2008}.  It should be remembered, however, that any observable will depend on integration over a range of dipole sizes, therefore even at high $Q^2$ there will be some contribution from large dipole sizes $r>r_S$.  Moreover, dipole models incorporating saturation fitted to HERA data may be extrapolated to very low $x$ and to predict cross sections for nuclear collisions where the saturation scale is enhanced by $A^{1/3}$ \cite{Kowalski:2007rw}.  In these situations, multi-Pomeron exchange may become important and extrapolation based on single-Pomeron exchange would be unreliable.

\section{Theory outlook: saturation beyond the BK equation in a statistical picture}

The BK equation describes unitarity corrections in the asymmetric configuration, when the target is extended and dense and the projectile is small and dilute.
 In a more symmetric situation, like  $\gamma^*(Q^2) p$ scattering at low 
$Q^2$, the BK approximation is no longer sufficient. In the diagrammatic formulation, besides the fan diagram one should then take into account diagrams with closed Pomeron loops.  To construct a fully reliable and practical theoretical treatment of this complex case has turned out to be a prohibitively difficult task so far. Fortunately, the key properties of solutions of the BK equation in the low momentum region follow from its universal features and do not rely on the details of the equation.

In the Kovchegov derivation of the BK equation \cite{Balitsky:1995ub,*Kovchegov:1999yj,*Kovchegov:1999ua} one uses the Mueller dipole cascade picture~\cite{Mueller:1993rr} of the small~$x$ QCD evolution. The equation expressed in terms of the dipole scattering amplitude, 
$N_{\ubo\vbo}(Y) \equiv {\cal N}(x,\boldsymbol{r},\boldsymbol{b})$, with
$Y=\ln(1/x)$, reads
\be
\label{eq:BK}
\frac{\partial N_{\ubo\vbo}}{\partial Y}=\frac{\alfasbar}{2\pi}\int \mathrm{d}^2\wbo\;
\frac{(\ubo-\vbo)^2}{(\ubo-\wbo)^2(\wbo-\vbo)^2}
\left[ N_{\ubo\wbo}+N_{\wbo\vbo}-N_{\ubo\vbo}-N_{\ubo\wbo}N_{\wbo\vbo}\right]\,
\ee
where 
$\boldsymbol{u} = \boldsymbol{b}  - \boldsymbol{r}/2$, and
$\boldsymbol{v} = \boldsymbol{b}  + \boldsymbol{r}/2$ 
(assuming $z$=1/2 in the definition of $\boldsymbol{b}$, cf.~Sec.~2).
The equation has two fixed points: the repulsive one, 
$N_{\ubo\vbo}=0$, from which the solution is driven out by the linear term,
and the attractive one, $N_{\ubo\vbo}=1$, where the linear and  
nonlinear term compensate each other. This scenario of linear growth
of the amplitude tamed by non-linear rescattering effects is  
common to all existing approaches to the saturation phenomenon.
In the uniform case, when $N$ does not depend on the impact parameter, 
$b$, this combination of growth and nonlinearity was shown to 
lead to a {\em geometric scaling} property~\cite{Stasto:2000er} of the 
solutions,
$
N_{\ubo\vbo}(Y)=N(|\boldsymbol{u}-\boldsymbol{v}|^2Q_s^2(Y)) \quad 
\mathrm{for} \;\; Y\gg 1,
$
irrespective of the initial conditions~\cite{Munier:2003vc,*Munier:2003sj,*Munier:2004xu}.
For the $\gamma^*p$ cross section, geometric scaling implies that 
$\sigma^{\gamma^*p}(x,Q^2)=\sigma^{\gamma^* p}(Q_s^2/Q^2)$, which was 
observed in HERA data \cite{Stasto:2000er}.

Interestingly enough, the geometric scaling property of the BK equation does not depend on the details of either the linear or the non-linear term. Therefore the scaling is a robust and universal phenomenon.
In particular, the BK equation belongs to the same universality class as a simpler and well understood Fisher--Kolmogorov--Petrovsky--Piscounov (FKPP) equation \cite{Munier:2003vc,*Munier:2003sj,*Munier:2004xu},  $\partial_t u(x,t)=\partial^2_{xx}u+u-u^2,$
where the rapidity is mapped onto the~time $t$ and the logarithm of the dipole
 size onto the real variable~$x$. Employing this connection it was proved 
that, indeed, both the emergence of geometric scaling and the rapidity 
evolution of the saturation scale are universal phenomena and do not 
depend on the details of the BK equation, provided that the initial condition 
is uniform in the impact parameter space.

The statistical framework implied by the Mueller dipole model may also be
used to provide some qualitative insight into the effect of ``Pomeron loops''
in the scattering amplitudes~\cite{Mueller:2004se,Munier:2005re,*Iancu:2004es}. This effect corresponds to a stochastic term added to the FKPP 
equation~\cite{Munier:2005re,*Iancu:2004es},
\be
\partial_t u(x,t)=\partial^2_{xx}u+u-u^2 + \sqrt{u(1-u)}\,\eta\,
\ee
where $\eta$ is the white noise. The origin of stochasticity can be traced 
back to the discreteness of the colour dipoles in the Mueller cascade model.
The BK equation is derived in the mean field approximation when 
the density of colour dipoles in the projectile is large enough ($n\gg 1$) 
that statistical fluctuations in the number of dipoles can be neglected. 
In this case, $N_{\ubo\vbo}$ is an averaged dipole scattering amplitude.
At the edge of the dense regime of the dipole distribution, however,  
the dipole occupation number is small, $n\sim 1$, so the statistical 
fluctuations play an important r\^{o}le. It was realised in Ref.~\cite{Mueller:2004se} 
and subsequently developed in Ref.~\cite{Munier:2005re,*Iancu:2004es}
that these fluctuations get enhanced in the $Y$-evolution and affect the 
global properties of the amplitude. In this approach the saturation scale
becomes a stochastic variable that fluctuates from one scattering event
to another, with a lognormal distribution with the variance $\sigma^2=D Y$,
where  $\,D \sim \alpha_s / \ln^3 (1/\alpha_s^2)\,$ 
\cite{Marquet:2006xm,*Enberg:2005cb}. 
The most important result of fluctuations is a new scaling of the physical amplitude, called {\it diffusive scaling} \cite{Munier:2005re,*Iancu:2004es}. 
Namely, the dipole scattering amplitude $N_{\ubo\vbo}(Y)$,  should depend only on one variable, $\,\xi =  (\ln(r^2)+\left<\ln Q_s^2\right>) / \sqrt{D Y}$.
Note that the factor $\sqrt{D Y}$ in the denominator which spoils the 
geometric scaling is of the diffusive origin. 
A first attempt to trace the diffusive scaling in the HERA data on 
$F_2$ was presented in Ref.~\cite{Basso:2008re} with a negative result. 
This would suggest that Pomeron loops introduce only a small effect 
in the HERA data.

The results presented here neglect the impact parameter dependence of the scattering amplitudes, assuming that the high energy QCD evolution is local in the transverse coordinate space. Thus the local evolutions at different $b$'s are uncorrelated. Recent numerical studies \cite{Munier:2008cg} suggest that this is a quite accurate picture of high energy scattering if the dipole size is significantly smaller than the target size.

Recently, an interesting attempt was made~\cite{Avsar:2005iz,*Avsar:2006jy,*Avsar:2007xg,*Flensburg:2008ag,*Avsar:2007} to explicitly model 
the colour dipole cascade taking into account effects 
related to Pomeron loops. In more detail, subleading effects in the 
$1/N_c$ expansion were phenomenologically incorporated  that lead to a 
possibility of colour dipole reconnections in the dipole wave function. 
The resulting dipole--dipole scattering amplitudes 
were shown to respect with good accuracy the symmetry between the target and 
the projectile, which does not hold in the absence of the colour reconnection. 
The approach employs Monte-Carlo methods and was shown to be quite successful 
in describing total cross-sections and many diffractive observables.

\section{Concluding remarks}

The dipole models applied to HERA data on inclusive and diffractive 
processes provide a successful unified description of most observables.
These analyses provide significant evidence for sizable unitarity 
(rescattering) corrections to the single scattering approximation, 
that is used in the linear QCD evolution equations, in both DGLAP and BFKL. 
These corrections become strong below the saturation scale, $Q_S(x)$. 
The determination of the saturation scale within different 
dipole models yields consistently that $Q_S < 1$~GeV, over the HERA kinematic 
range. $Q_S$ is found to increase with~$1/x$, approximately as 
$Q_S^2(x) \sim (1/x)^{\lambda_S}$ with $\lambda_S \simeq 0.12\,$--$\,0.2$, 
depending on the model. Both these properties of~$Q_S$ suggest 
that the onset of perturbative saturation is probed at HERA, 
and that non-perturbative effects may still be significant around $Q_S$. 
Fortunately, the key results on the saturation phenomenon obtained within
perturbative QCD are universal and should remain valid despite a possible 
non-perturbative contamination.

\section*{Acknowledgements}
K.~G.-B.\ and L.~M. acknowledge a support of grant of Polish Ministry of 
Education No.\ N N202 249235. L.~M.\ is supported by the DFG grant SFB~676.  
G.~W.\ acknowledges the UK Science and Technology Facilities Council for the    award of a Responsive Research Associate position. 

\bibliographystyle{heralhc} 
{\raggedright
\bibliography{heralhc}
}

\end{document}